\documentclass[twocolumn,aps,prb,groupedaddress,showpacs,amsmath,amssymb]{revtex4}

\renewcommand{\Im}{\textrm{Im}}

\newcommand{\eq}{\textrm{equation}}
\newcommand{\Eq}{\textrm{Equation}}
\newcommand{\mean}[1]{{\langle #1\rangle}}

\newcommand{\tl}[1]{_{\textrm{#1}}}

\usepackage{graphicx}
\usepackage{bbm}
\usepackage{psfig}
\usepackage{epsfig}
\usepackage{dcolumn}
\usepackage{bm}

\begin{document}
\title{Interface steps in field effect devices}
\author{Samuel Wehrli}
\email{swehrli@phys.ethz.ch}
\author{Christian Helm}
\affiliation{Theoretische Physik, ETH-H\"onggerberg, CH-8093 Z\"urich, 
             Switzerland}
\date{\today}

\begin{abstract}
The charge doped into a semiconductor in a field effect transistor
(FET) is generally confined to the interface of the semiconductor. 
A planar step at the interface causes a potential drop due to the strong 
electric field of the FET, which in turn is screened by the doped carriers. 
We analyze the dipolar electronic structure of a single step 
in the Thomas-Fermi approximation and find that the transmission
coefficient through the step is exponentially suppressed by the
electric field and the induced carrier density as well as by the step height.
In addition, the field enhancement at the step edge can facilitate the
electric breakthrough of the insulating layer.
We suggest that these two effects may lead to
severe problems when engineering FET devices with very high doping. 
On the other hand steps can
give rise to interesting physics in superconducting FETs by forming
 weak links and potentially creating atomic size Josephson junctions. 
\end{abstract}


\maketitle

\section{Introduction}

\begin{figure}
\begin{center} 
\begin{minipage}[b]{0.45\textwidth}
\includegraphics[width=\textwidth]{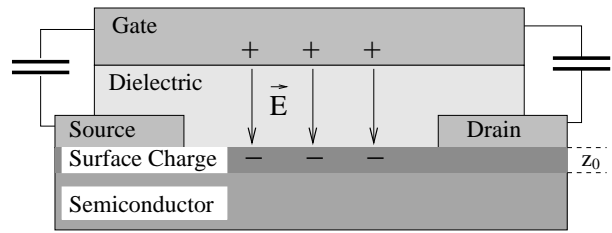}
\end{minipage}\vspace{0cm}\\
\end{center}
\caption{\label{fig:fet}
Schematic picture of field effect transistor (FET). }
\end{figure}

The field effect transistor is a widely used 
device, both for commercial products as well as in research on correlated
electron system (a recent overview is
given in Ref.~\onlinecite{ahn}). 
Traditionally this technique
is applied to semiconductors, such as silicon or GaAs, and more recently to
high temperature superconductors (see Ref.~\onlinecite{ahn,mannhart} 
and references
therein) as well as 
organic materials~\cite{yang,lin, butko, podzorov,deboer, takeya}. The
field effect is intriguing because it allows in principle to tune
 the charge density continuously by changing 
the applied voltage between gate and source/drain (see
Fig.~\ref{fig:fet}). However, field effect 
experiments on the systems mentioned above 
are difficult, because high electric fields, limited by the
electric breakdown of the insulating dielectric,  are required in
order to achieve substantial changes of the charge distribution.  
Nevertheless,
new gate insulator materials, such as complex oxide dielectrics and
ferroelectric oxides,  in recent years allowed to push induced surface
charge densities to promising values~\cite{ahn}. 
Thereby operating a FET at high  electric fields leads 
 to a strong confinement of the induced surface charge to the
dielectric-semiconductor interface. Therefore, imperfections of the interface
become important. 

In the present paper 
the effect of interface steps in such devices is considered. Steps occur
as imperfections when growing crystals~\cite{lin}, or can be created
artificially. Naturally, one expects that steps act as barriers in the
transport channel confined to the surface. Below we show under what 
circumstances this
effect should be important, and that the transmission through the step 
depends exponentially on the electric field and the step height. 
Therefore, at high enough doping, a step cuts the transport channel and may
present a severe problem. This may be particularly important  
in organic FETs where the step height is given be the molecular size and
therefore is quite big. However, in superconducting FETs the step-barrier
can act as a weak link and give rise to a  Josephson junction where the
critical current depends exponentially on the applied field.
This might be an interesting way to design 
a dissipationfree switch in a superconductor~\cite{gianni}. 

We distinguish two types of field effect transistors.
The term metal-insulator-semiconductor FET (MISFET) is used
for a whole class of devices where the intrinsic 
carrier density of the semiconductor is negligible
($N_i\approx 0$) in the absence of the electric field.
The electric field serves to induce a conducting space-charge layer on
the surface (see Fig.~\ref{fig:fet}). In contrast to this, in the
superconducting FET (SUFET), the semiconductor is replaced by a superconductor
or metal (above $T_c$) 
which has a finite density $N_i$. In this case the electric field  
alters the density which, for example, may change the superconducting
$T_c$. In both cases the electric field induces a surface charge
which screens the field. However, 
the extension $z_0$ of the surface charge perpendicular to the
interface differs for MISFET and SUFET due to the difference in $N_i$. As
a surface step of height $h$ is expected to be important for $h\gtrsim z_0$, 
we first give a rough estimate of $z_0$ for both cases.

The problem of the charge profile in MISFETs was extensively studied in
Ref.~\onlinecite{ando} for the case of a continuous medium. Calculations for
molecular crystals, which explicitly take into account the
discreteness of the lattice, were done recently~\cite{sam,sinova}. 
Below we follow Ref.~\onlinecite{ando} and refer to this work for more details.
In the case of a
perfectly flat and infinite 
interface, the wavefunction of the
carriers can be separated into a plane-wave part parallel to the interface
and a transverse part $\zeta(z)$, where the $z$-direction is perpendicular to
the interface. The charge profile is then given by
$n(z)\propto ¦ \zeta(z)¦^2 $. 
The transverse function $\zeta$ is best calculated using a trial
wavefunction for the lowest subband, such as the Fang-Howard trial wavefunction
$\zeta(z)\propto z\exp(-bz/2)$, where $b$ is the variational parameter. 
We use this ansatz which is sufficient for
our purpose, although more accurate trial functions exist~\cite{morf}. 
Minimizing the energy with respect to $b$ yields the
average distance $z_0$ 
of the charge distribution from the interface and the width $w$ of
the distribution:
\begin{eqnarray} \label{eq:z0}
   z_0 &=& \langle z \rangle=\frac{3}{b}=
   \left(\frac{18\,\varepsilon_s\,\hbar^2}{11\,\pi\, m_z\, e^2\, n_0}\right)^{\frac{1}{3}},\\
   w   &=& \sqrt{\langle z^2\rangle-
     \langle z \rangle^2}=\frac{z_0}{\sqrt 3}.
\end{eqnarray}
In the equation above $\varepsilon_s$ is the dielectric constant of the
semiconductor, $m_z$ its effective mass in the $z$-direction 
and $n_0$ the induced surface carrier
density. Note that we use a small $n$ for surface density and a capital
$N$ for the volume density. 
Estimates of $z_0$ are given in Tab.~\ref{tab:parameters}. Parameters for the
Si-FET are standard and given in the literature. At the highest possible
fields, $z_0=12$~\AA{}, which corresponds to  a few unit cells. 
Pentacene was chosen as an
example for an organic MISFET, and the parameters for the thin film transistor
presented in Ref.~\onlinecite{yang} were taken. Only the effective mass $m_z$
is not 
accessible by experiment and has to be 
estimated from band-structure calculations.
Theoretical calculations
 yield a hopping  $t_z=0.47$~meV between the Pentacene molecules in
the $z$-direction~\cite{cornil}. Assuming a tight-binding 
band, the effective mass
at the minima of the band is given by $m_z=\hbar^2/(2\,a^2\,t_z)$, where
$a=15.5$~\AA{} is the layer spacing in the $z$-direction~\cite{lin}. 
This yields
$m_z=0.34$~$m_0$. \Eq~(\ref{eq:z0}) gives then $z_0=20$~\AA{} which is of
the same order as the layer spacing. Hence, molecular steps, as are observed
when growing films, may be important in such devices~\cite{lin}. Note that 
up to now, pentacene samples used in the experiments have many traps which may
lead to corrections to the estimation of $z_0$ in Eq.~(\ref{eq:z0}). 
However, there are experimental attempts to use single crystals in order to
reduce the number of traps~\cite{butko}. 

In the case of SUFETs the superconductor has a finite density $N_i$ of
carriers which give raise to metallic screening. Electric fields are
screened within the Thomas-Fermi screening length $\lambda\tl{TF}$. Therefore 
$z_0\approx\lambda\tl{TF}$ for SUFETs.
In cuprates (e.g. YBa$_2$Cu$_3$O$_7$)
$N_i\approx 1$-$5\times 10^{21}$~cm$^{-3}$ and the Thomas-Fermi
screening length is estimated to be 5-10~\AA{}~ (Ref. \onlinecite{mannhart}).
This is larger than the width of the superconducting layer~(3~\AA{}), but
smaller then the unit cell ($a=15$~\AA{}), which means that only the first
superconducting layer is affected by the electric field. Therefore, it is
important that the transport in the SUFET occurs only in the first layer.
Note that in a SUFET the drain and source electrodes usually contact 
 the first layer (see Fig.~\ref{fig:fet}). 
However, the Josephson coupling among layers leads to a shortcut through the
bulk. There are different possibilities to resolve this problem: (a) one uses
a very thin superconductor which, ideally, is only one unit cell
thick~\cite{ahn2}. (b) 
The distance between superconducting layers is increased in order to
suppress the Josephson coupling~\cite{tafuri}. 
(c) The interlayer Josephson coupling is suppressed e.g. by a parallel 
magnetic field (Fraunhofer like pattern).
In Ref.~\onlinecite{ahn2} possibility (a)  was realized and 
changes in $T_c$ as well as an insulating phase were
induced by the field effect. The ferroelectric Pb(Zr$_x$Ti$_{1-x}$)O$_3$ was
used as a gate dielectric and a surface carrier density 
$n_0=7\times 10^{13}$~cm$^{-2}$
was achieved. The superconductor consisted of  ~1 to 2 unit cells of
GdBa$_2$Cu$_3$O$_{7-x}$. Similarly, a single superconducting
CuO$_2$-layer has been created due to inhomogeneous oxygen doping
(resulting in a distribution of $T_c$ in the layers) 
on the surface of a Bi$_2$Sr$_2$CaCu$_2$O$_8$ single crystal  in Ref.~
\onlinecite{moessle}.  
Again, in this device the physics is confined to a
small region at the interface and steps should play an important role. In
particular, when the sample is superconducting, a step may induce a weak link.
This effect can be of particular relevance in highly anisotropic layered superconductors
such as the high-$T_c$ material Bi$_2$Sr$_2$CaCu$_2$O$_8$
or intercalated (LaSe)$_{1.14}$NbSe$_2$, see Ref. \onlinecite{samuely}, 
where the thickness/coherence length 
$\xi_\perp \approx 3$~\AA{} of the superconducting (CuO$_2$ or
NbSe$_2$)-layers  is much smaller than their distance ($d \approx 15$~\AA{}), 
i.e. atomic-size steps can naturally constitute a weak link (case b). 

In the following we solve first the problem of an interface step of a perfect
conductor exposed to a perpendicular electric field. This is a good
approximation for $h\gg z_0$ and yields an induced dipole. Knowing  the charge
dipole, the potential barrier across the step is calculated using the
Thomas-Fermi approximation.
 Finally, the transmission coefficient for
tunneling through the step is calculated in the WKB approximation.
%
%
\begin{table}
\begin{tabular}{l||c|c||c}
     & MISFET & MISFET & SUFET \\
     & Si$^a$ & Pentacene$^b$ &  GdBa$_2$Cu$_3$O$_{7-x}$$^c$  \\
  \hline
  Dielectric & SiO$_2$ & SiO$_2$ & Pb(Zr$_x$Ti$_{1-x}$)O$_3$ \\
  $\varepsilon\tl d$ & 3.9$^a$ & 3.9$^a$  & $-$ \\
  $\varepsilon\tl s$ & 11.5$^d$ & 6.7$^b$ & $-$  \\
  $m_z$ ($m_e$) & 0.916$^d$ & ~0.34$^e$  & $-$ \\
  $N_i$ (cm$^{-3}$) & 0 & 0  & $10^{21}$ \\
  \hline
  $E_b$ (MV/cm)& 10 & 3  & $-$ \\
  $n_0$ (cm$^{-2}$) & $2\times 10^{13}$ & $7\times 10^{12}$& 
              $-7\times 10^{13}$ \\
  $z_0$ (\AA{}) & 12 & 20 & $5-10^f$ \\
  \hline
  $h$ (\AA{}) & 20 & 15.5 & 12 \\
  $n_i$ (cm$^{-2}$) & 0 & 0 & $1.2\times 10^{14}$ \\
  $1/k_F$ (\AA{})  & 9 & 15 & 5.6 \\
  $\mean T$ & 0.06 & 0.2 &  0.02
\end{tabular} 
\caption{\label{tab:parameters}
  Comparison between different FETs. 
  $\varepsilon_d$ and $\varepsilon_s$ 
  are the dielectric constants of the dielectric and the
  semiconductor respectively.
  $m_z$ is the effective mass in the $z$-direction 
  which enters \eq~(\ref{eq:z0}). 
  $N_i$ is the intrinsic carrier density,
  $E_b$ is the sample dependent breakdown field and
  $n_0=E_b\varepsilon_d/(4\pi\,e)$ is the maximum
  surface carrier density.
  $z_0$ is the distance of the charge distribution from the interface, which
  is calculated by Eq.~(\ref{eq:z0}) for the MISFET and
  equal to the Thomas-Fermi screening length $\lambda\tl{TF}$
  in the case of the SUFET.
  $h$ is the step height which, for Pentacene and GdBa$_2$Cu$_3$O$_{7-x}$,
  was taken to be the layer spacing in the $z$-direction.
  $n_i=N_i h$ is the intrinsic surface carrier density in a single layer.
  $k_F=\sqrt{2\pi(n_i+n_0)}$ 
  is the Fermi-wave vector.
  $\mean T$ is the average transmission coefficient calculated by
  \eq~(\ref{eq:meanT}). 
  $^a$Ref.~\onlinecite{ahn},
  $^b$Ref.~\onlinecite{yang},
  $^c$Ref.~\onlinecite{ahn2},
  $^d$Ref.~\onlinecite{ando},
  $^e$See text,
  $^f$Ref.~\onlinecite{mannhart}.
}
\end{table}
%

\section{perfect conductor approximation}

\begin{figure}
\begin{center} 
\begin{minipage}[b]{0.4\textwidth}
\includegraphics[width=\textwidth]{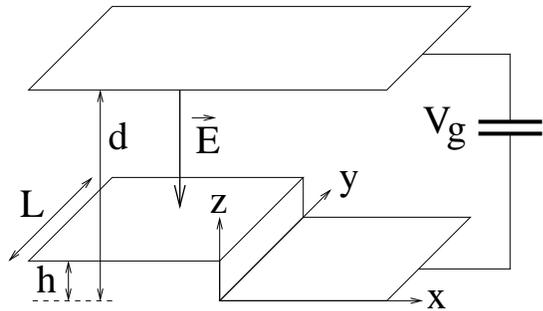}
\end{minipage}\vspace{0cm}\\
\end{center}
\caption{Geometry of an interface step. 
\label{step}
}
\end{figure}

\begin{figure}
\begin{center} 
\epsfig{file=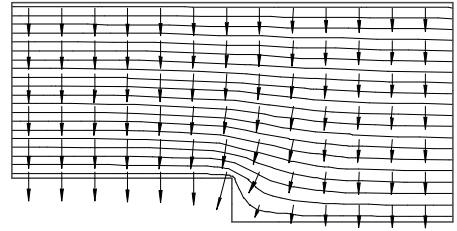,width=0.3\textwidth,clip=,angle=-90}
\end{center}
\caption{Equipotential curves and electric field vectors near a step
as in Fig.~\ref{step}.\label{equipot}
}
\end{figure}

\begin{figure}
\begin{center}
\begin{minipage}[b]{0.4\textwidth}
  \begin{center}
    \includegraphics[width=\textwidth,clip=true]{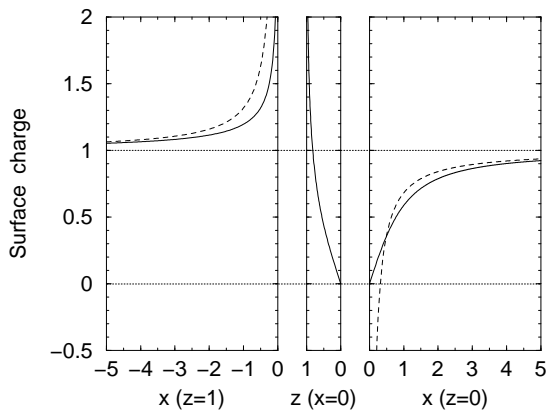}
  \end{center}
\end{minipage}\vspace{0cm}\\
\caption{\label{fig:CondStep}
Surface charge on the perfect conductor (in units of $n_0$) along the
step (distances in units of the step height $h$). \emph{Solid line:} Exact
solution with the divergences $x^{-1/3}$ or $(h-z)^{-1/3}$ at
(x,z)=(0,h). 
\emph{Dashed line:} Approximate solution~(\ref{eq:napprox}) with the
$\sim 1/x$ divergence. 
}
\end{center}
\end{figure}

In the geometry with a step of height $h$  along the $y$-direction  (see
Fig.~\ref{step}) we  model the interface by a 
two-dimensional grounded conductor with a potential difference $V_{\rm G}$ 
to the gate in distance $d$.
This approximation becomes exact, if the thickness of the
charged area is small compared to the step height, $ z_0 \ll h$. 
Far from the step the electric field is 
homogenous, $E_0 = V_{\rm G} / d \approx V_{\rm G} / (d+h)$,
 but it is distorted near the step, as a surface 
charge is induced there in order to compensate 
the potential difference $\Phi_0 = e V_{\rm G} h  / d$, which is 
realized for a flat interface with a homogeneous charge distribution. 
The typical equipotential curves near the step are shown in
Fig.(\ref{equipot}). 
As the system is invariant in the $y$-direction, we can use the conformal
transformation (in the $xz$-plane) 
\begin{equation} 
\label{eq:map}
  u=x+iz=
  \frac{h}{\pi}\left[\sqrt{w^2-1}+\log\left(w+\sqrt{w^2-1}\right)\right], 
\end{equation}
which maps the upper half complex 
plane onto the domain above the step, the cuts
being on the positive real axis. 
The solution for the potential is then given by     
\begin{equation} \label{eq:pot}
  \Phi(x,z)=-\frac{E_0\,h}{\pi}\, \Im [w(u=x+iz)] . 
\end{equation}
Far from the step, $w(u)$ from \eq~(\ref{eq:map}) can be expanded in orders of
$h/\sqrt{x^2+z^2}$,  which yields the approximate expression 
\begin{equation} \label{eq:potapprox}
  \Phi(x,z)=-E_0\left(z-h\;\frac{\theta(x,z)}{\pi}\right), 
  \quad \sqrt{x^2+z^2}\gg h\; , 
\end{equation}
where $\theta$ is the polar angle in the $xz$-plane.
In general, the surface charge of a perfect conductor is given by 
$n=\varepsilon_{d} E_\perp/(4\pi e)$, where $E_\perp$ is the electric field being
perpendicular to the interface and $\varepsilon_{d}$ is the dielectric
constant of the gate insulator. Using \eq~(\ref{eq:potapprox}) for
the potential yields the asymptotic charge distribution ($x\gg h$)
\begin{equation} \label{eq:napprox}
  n(x)=n_0\left(1-\frac{1}{\pi}\,\frac{h}{x}\right) , 
\end{equation}
with the doping  $n_0=\varepsilon_{d} E_0/(4\pi e)$. The 
exact solution is given by the parametric expression 
\begin{equation} \label{eq:nexact}
  n(x,z)=n_0\left|\frac{1-w(u)}{1+w(u)}\right|^{\frac{1}{2}} , 
\end{equation}
where $w(u)$ is given by Eq.~(\ref{eq:map}) and $u=x+iz$ is taken on
the surface of the conductor. 
The exact result as well as the approximation 
\eq~(\ref{eq:napprox}) are shown in Fig~\ref{fig:CondStep}.
Obviously,  the electric field induces a dipole center at the step 
which falls off as
$x^{-1}$ far from the step.

It can be shown by the exact solution~(\ref{eq:nexact}) that there is a weak 
divergence ($\sim x^{-1/3}$ or $\sim (h-z)^{-1/3}$ respectively)
of the surface charge at the upper corner of the 
step due to the sharp edge, which enhances the local tunneling
rate through the dielectric and can therefore serve as a nucleation center
for a possible breakthrough of the device.  
The singularity at the edge can be regularized: 
(a) 
at distances 
from the step smaller than the local 
$\lambda_{\rm TF}$, where the quantum mechanical 
exclusion principle comes into play, 
(b)
geometrically due to finite step curvature. In both cases
the field enhancement is of the order $(h/\lambda_c)^{1/3}$ where 
$\lambda_c$ is the cutoff ($\lambda_{\rm TF}$ or finite curvature) which is at
least of the order of 
the atomic scale ($\lambda_c\approx 1-2$~\AA{}). 
Hence, ratios $h/\lambda_c$ might
be of the order $10-20$ and the field enhancement is expected to be a
factor $2-3$. This is not a big factor, but it should nevertheless 
be important, because
steps are line defects which have a big probability to hit a ``weak spot'' in
the dielectric where an electric breakthrough can occur. Note also that
kinks in the step would lead to a further enhancement of the electric
field. 
(c) 
The singularity can also be regularized
due to the finite bandwidth, where the locally induced charge exceeds the 
filling of the band. This occurs when the potential difference  
$\Phi_0 = e V_{\rm G} h / d$ exceeds the bandwidth. For the maximal
gate fields $\Phi_0 = E_b \, h$ which is 2~eV for the
Si-MISFET and 0.5~eV for the pentacene MISFET
(see Tab.~\ref{tab:parameters}). In the latter case this is roughly the
bandwidth and    
a complete local charge depletion can be expected~\cite{cornil}.

\section{Transmission across the step}

As discussed in the introduction, the wavefunctions of the electrons confined
to a flat and infinite interface can be separated into a parallel and a 
transverse part: 
$\psi_{{\bf k}j}({\bf r},z)=e^{i{\bf kr}}\zeta_j(z)$  
where ${\bf r}=(x,z)$ and ${\bf k}=(k_x,k_z)$ represent
coordinate and wavevector parallel to the interface and $j=0,1,2\ldots$ labels
the discrete and confined states in the transverse direction. 
The energies are given by 
\begin{equation}
  \epsilon_{{\bf k}j}=\frac{\hbar^2 k^2}{2 m_{\parallel}}+\epsilon_j.
\end{equation} 
It can be shown that the eigenvalues $\epsilon_j$ of the tranverse part
scale as $\epsilon_j\propto E_0^{2/3}\propto n_0^{2/3}$, where $E_0$
and $n_0=\varepsilon_d E_0/(4\pi e)$ are the electric field and the surface
carrier 
density respectively (see Ref.~\onlinecite{ando}).
If only the $j=0$ state is occupied (i.e. the lowest subband), 
then the system is two-dimensional and the  Fermi
energy is given by $\epsilon_F=n_0/g$, where $g=m_{\parallel}/(\pi\,\hbar^2)$
is 
the constant density of states. The condition for the occupation of only the
lowest subband is $\epsilon_F < \epsilon_1-\epsilon_0$ which holds for 
densities $n_0$ below a certain threshold. in
Si, this threshold is $3\times10^{13}$~cm$^{-2}$ (using the triangular
potential approximation, see Ref.~\onlinecite{ando}) which is of
the same order as the maximally achievable surface densities 
(see Tab.~\ref{tab:parameters}). Hence, 
it is not clear, whether only the lowest
subband is occupied. However, the
presence of a step leads to charge dipole which suppresses the carrier
density on the lower side of the step where the
condition is then clearly satisfied. As
discussed below, it is exactly this region which is 
of interest when calculating
the transmission coefficient through the step. 
Therefore, we consider in the following only the lowest subband and treat
the system as two-dimensional.
The non-uniform charge distribution, caused by the step,  can be
treated in the Thomas-Fermi approximation (TFA) which locally assumes a free
(here 2D) electron gas with density 
\begin{equation}\label{eq:tfa}
  n({\bf r})=g \epsilon_F({\bf r})=g[\epsilon_F-V\tl{loc}({\bf r})],  
\end{equation}
where $\epsilon_F({\bf r})$ is the spatially dependent Fermi Energy.
At this point the result of the previous section enters by
assuming that the surface carrier density 
$n({\bf r})=\int dz\, N({\bf r},z)$ 
of the real system 
is still well approximated
by the perfect conductor approximation (PCA). Note that in the PCA 
the induced charge is a pure surface charge with zero width and therefore
differs from the real charge distribution which has a finite extension in the
$z$-direction.
However, the PCA becomes exact far from the step.
Furthermore, knowing the solution of the PCA allows to calculate the local
potential $V\tl{loc}$ via relation~(\ref{eq:tfa}).
In the following the approximate solution~(\ref{eq:napprox}) of the PCA is used
which has the right asymptotics and which permits an analytic calculation of
the transmission coefficient.
Combining relations~(\ref{eq:napprox}) and~(\ref{eq:tfa}) yields 
\begin{equation} \label{eq:tfapot}
  V\tl{loc}(x)=\frac{1}{\pi}\,\frac{n_0}{g}\,\frac{h}{x}
      =\frac{\hbar^2\,n_0}{m_{\parallel}}\,\frac{h}{x}.
\end{equation}
The potential~(\ref{eq:tfapot})
acts as a potential barrier across the step for $x>0$.
In the WKB approximation the transmission coefficient $T({\bf k})$ of a mode 
${\bf k}$ is given by
\begin{equation} 
  T({\bf k})=\exp\left(  
    -2\int_0^{x_0}\sqrt{(2m_{\parallel}/h^2)
    [V\tl{loc}(x)-\tilde\epsilon_{\bf k}]}\,dx  \right),
\end{equation}
where 
$\tilde\epsilon_{\bf k}=
\epsilon_{{\bf k}0}-\epsilon_0-\hbar^2\,k_y^2/(2m_{\parallel})=
\hbar^2\,k_x^2/(2m_{\parallel})$ is the kinetic energy along the
$x$-direction.  
Using the local potential~(\ref{eq:tfapot}) implies
$x_0=n_0 h /(\pi g \epsilon)=2n_0h/k_x^2$  and yields the analytic result 
\begin{equation} \label{eq:T}
  T({\bf k})=\exp\left(-2\pi n_0\frac{h}{|k_x|}\right).
\end{equation}
Not surprisingly, the transmission coefficient depends exponentially
on the step height $h$ and on the electric field which is proportional to the
surface density $n_0$. In order to calculate the conductance through the
step, the average $\mean{T}$ over the Fermi surface of the transmission
coefficient enters, which for the parabolic band is  
\begin{equation} \label{eq:meanT}
  \mean{T}=\int_{0}^{2\pi}\frac{d\varphi}{2\pi}\,T(k_F\cos\varphi)
          =f\left(2\pi n_0\frac{h}{k_F}\right), 
\end{equation}
where the function $f$ is given by
\begin{equation} \label{eq:f}
  f(x)=\frac{2}{\pi}\int_{0}^{\pi/2}d\varphi\,e^{-x/\cos\varphi}
     \approx\sqrt{\frac{2}{\pi}}\;\frac{e^{-x}}{\sqrt x}.
\end{equation}
The approximate expression is valid for $x\gg 1$. 
In a 2D system with parabolic dispersion the Fermi wave-vector is 
$k_F=\sqrt{2\pi (n_i+n_0)}$, where $n_i$ is the intrinsic
carrier density. In the MISFET $n_i=0$ which yields the simple expression
$\mean{T}=f(k_F h)$.
The conductance through the
step is given by the Landauer Formula 
\begin{equation} \label{eq:G}
  G = G_0\,N\,\mean T,
\end{equation}
where $G_0=2e^2/h\approx 12.9$~k$\Omega$ is the quantum conductance. 
$N=(2/\pi)\,k_F L$ is the number of channels for a sample of size $L$ 
(along the $y$-direction) taking into account both spins and the condition 
$-k_F<k_y<k_F$. The functional behavior of
$f$ implies that the conductance through steps with height
$h>k_F/(2\pi n_0)$ (in the MISFET $h>1/k_F$) are exponentially suppressed. 

\Eq~(\ref{eq:meanT}) was applied to the FET examples discussed previously and
the results are summarized in Tab~\ref{tab:parameters}. An arbitrary step
height $h=20$~\AA{}, which might be due to an artificial step, 
was taken for the Si-MISFET, which yields $\mean{T}=0.06$ for the highest
fields. In the Pentacene MISFET where $1/k_F=15$~\AA{} which is almost 
the same as the interlayer spacing ($a=15.5$~\AA{})
and which yields $\mean{T}=0.2$ for a molecular step ($h=a$) at the highest
achievable dopings. 

Recently, there were unconfirmed 
claims that pentacene would become superconducting when
doped to half filling~\cite{beasley}. In the light of our discussion such
high dopings are unlikely, as the locally enhanced electric
field at the step edges might lead to the
electric breakthrough in 
the dielectric. Even if such electric
fields could be reached in reality, the resulting device would be
significantly limited by molecular steps, as for 
required density $n_0=2\times 10^{14}$~cm$^{-2}$ corresponding to 
$1/k_F=3$~\AA{} the transmission $\mean{T}=2\times 10^{-3}$ of the
step would be small.

In lower electric fields, for the  SUFET below $T_c$, 
the charge dipole due to the step can form a
Josephson junction, which is of the SIS or SNS-type depending on the charge
density in the nonsuperconduting region. The critical current $I_c$ through
this junction can be estimated for a SNS tunnel junction by the Ambegaokar-Baratoff formula 
~\cite{ambegaokar}
\begin{equation}
  I_c R = I_c G^{-1} = 
   \frac{\pi\Delta}{2e}\,\tanh\left(\frac{\Delta}{2kT}\right),
\end{equation}
where $\Delta$ is the ($s$-wave) gap of the superconductor and $G$ the conductivity from
\eq~(\ref{eq:G}). In this case the critical current $I_c$ depends on the external
electric field exponentially via the effective thickness of the insulating
region, which enters the tunnel matrix element $T$. A similar exponential
dependence $I_c \sim \exp ( - l/l_T) $ is found in field effect doped
SNS-junctions with a large distance $l$ of the superconducting leads, 
$l \gg l_T = (\hbar D/ 2 \pi k_B T)^(1/2)$ ($D\sim n^{1/2}$ Diffusion constant in normal
metal), while in the opposite limit $l \ll l_T$ the critical current depends
algebraically on $l$, $I_c \sim 1/l$, see Ref. \onlinecite{volkov}. In both
cases it is seen that the superconducing transport depends sensitively on the
local charge density near the step, i.e. can be easily modified by an external
electric field. 

Note that in a highly anisotropic layered superconductors, such as
 high-$T_c$-materials an atomic size step can already form a
weak link in the absence of the charge dipole effect described here, 
because the coherence length ($\xi=3$~\AA{}) of the quasi 2D superconducting layers 
is smaller than the height $h$ of the step.

To conclude, we showed that a charge dipole is induced in the 2D electron 
gas near an interface step, which can lead to a (Josephson) weak link due to 
the local depletion of the charge density. This forms the limiting factor
for transport through an ultrathin metallic or superconducting layer and might
be used for ultrasmall dissipationfree switches (SUFET). In addition to this,
the field enhancement near the step in the insulating barrier can
trigger the
breakthrough of the dielectric and thereby limit the maximal doping level
$n_{0,{\rm max}}$. Both effects are hard to avoid and pose a fundamental
challenge in term of atomically flat interfaces for any FET device with a quasi
2D (super)conducting charge density.

We thank M.~Sigrist, T.M.~Rice and G.~Blatter for useful discussions and  the 
NRCC MaNEP of the Swiss NSF for financial support. 


\begin{thebibliography}{999}
\bibitem{ahn} C.H. Ahn, J.-M. Triscone, J. Mannhart, 
  Nature {\bf 424}, (2003) 1015.
\bibitem{mannhart} J. Mannhart, Supercond. Sci. Technol. {\bf 9}, (1996) 49.
\bibitem{yang} Y.S. Yang, S.H. Kim, J.I. Lee, H.Y. Chu, L.M. Do, H. Lee, 
  J. Oh, T. Zyung, M.K. Ryu, M.S. Jang, 
  Appl. Phys. Lett. {\bf 80}, (2002) 1595. 
\bibitem{lin} Y.Y. Lin, D.J. Gundlach, S.F. Nelson, T.N. Jackson,
  IEEE Trans. Electron Devices {\bf 44}, (1997) 1325.
\bibitem{butko} V.Y. Butko, X. Chi, D.V. Lang, A.P. Ramirez,
  cond-mat/0305402 (2003);
  V.Y. Butko, X.Chi, A.P.Ramirez, cond-mat/0307372 (2003).
\bibitem{podzorov} V. Podzorov, V.M. Pudalov, and M.E. Gershenson,
  Appl. Phys. Lett. {\bf 82}, (2003)  1739;
  V. Podzorov, S. E. Sysoev, E. Loginova, V.M. Pudalov,
  M.E. Gershenson, cond-mat/0306192 (2003). 
\bibitem{deboer} R.W.I. de Boer, A.F. Morpurgo, T.M. Klapwijk,
  cond-mat/0307320 (2003)
\bibitem{takeya} J. Takeya, C. Goldmann, S. Haas, K.P. Pernstich,
  B. Ketterer, B. Batlogg, cond-mat/0306206 (2003). 
\bibitem{switch} M.J. Storcz, F.K. Wilhelm, Appl. Phys. Lett. {\bf 83} (2003) 2387.  
\bibitem{gianni} L.B. Ioffe, V.B. Geshkenbein, M.V. Feigel'man, 
  A.L. Fauch\'ere, G. Blatter, Nature {\bf 398}, (1999) 679.
\bibitem{ando} T. And\'{o}, A.B. Fowler, F. Stern, Rev. Mod. Phys. {\bf 54},
  (1982) 437.
\bibitem{sam} S. Wehrli, D. Poilblanc, T.M. Rice, Eur. Phys. J. B {\bf 23}, 
  (2001) 345;
  S. Wehrli, E. Koch, M. Sigrist, to be published in PRB
\bibitem{sinova} J. Sinova, J. Schliemann, A.S. Nunez, A.H. MacDonald,
  Phys. Rev. Lett. {\bf 87}, (2001) 226802.
\bibitem{morf} R.H. Morf, N.  d'Ambrumenil, S. Das Sarma,
  Phys. Rev. B {\bf 66}, (2002) 075408.
\bibitem{cornil} J. Cornil, J.Ph. Calbert, J.L. Br\'edas,
  J. Am. Chem. Soc. {\bf 123}, (2001) 1250.
\bibitem{ahn2} C.H. Ahn, S. Gariglio, P. Paruch, T. Tybell, L. Antognazza, 
  J.M. Triscone, Science {\bf 284}, (1999) 1152
\bibitem{tafuri} F. Tafuri, J.R.~Kirtley, P.G.~Medaglia, P.~Origiani,
 G.~Balestrino, unpublished.
\bibitem{moessle} M.~M{\"o}ssle, R.~Kleiner, R.~Gatt, M.~Onellion,
  P.~M{\"u}ller, Physica C {\bf 341} (2000) 1571. 
\bibitem{samuely} P.~Samuely, P.~Szab\'{o}, J.~Kacmarcik, A.G.M.~Jansen,
  A.~Lafond, A.~Meerschaut, A.~Briggs, Physica C {\bf 369} (2002) 61. 
\bibitem{beasley}
 M.R.~Beasley (Chair), S.~Datta, H.~Kogelnik, H.~Kroemer, D.~Monroe:
 {\it Report of the Investigation Committee on the Possibility of Scientific
 Misconduct in the Work of Hendrik Sch\"on and Coauthors}, Sept.~2002,
 {\tt www.lucent.com/news\_events/pdf/researchreview.pdf} or
 {\tt publish.aps.org/reports/}.
\bibitem{ambegaokar} V. Ambegaokar, A. Baratoff, Phys. Rev. Lett. {\bf 10},
  (1963) 486
\bibitem{volkov} A.F. Volkov, H. Takayanagi, Phys. Rev. B {\bf 53} (1996)
  15162. 
\end{thebibliography}
\end{document}